\documentclass[journal]{IEEEtran}
\usepackage{amsthm,amsmath,amssymb}
\usepackage{dsfont}
\usepackage{mathrsfs}
\usepackage{amssymb,amsmath}
\usepackage{algorithmic}
\usepackage{algorithm}
\usepackage{url,subfigure,epsfig,graphicx}
\usepackage{float}
\usepackage{makecell}

\ifCLASSOPTIONcompsoc

  \usepackage[nocompress]{cite}
\else

  \usepackage{cite}
\fi

\ifCLASSINFOpdf

\else

\fi

\begin{document}
%

\title{Mobile Metaverse: A Road Map from Metaverse to Metavehicles}
%

\author{
\IEEEauthorblockN{Yilong Hui$^1$,~Gaosheng Zhao$^1$,~Nan Cheng$^1$,~Haibo Zhou$^2$,~and Zhou Su$^3$}
\\
%
\vspace{0.2cm}
\IEEEauthorblockA{$^1$State Key Laboratory of Integrated Services Networks, Xidian University, Xi'an, China\\
$^2$School of Electronic Science and Engineering, Nanjing
University, Nanjing, China\\
$^3$School of Cyber Science and Engineering, Xi’an Jiaotong University, Xi’an, China\\
}}


\IEEEtitleabstractindextext{
\begin{abstract}
With the rapid development of communication technologies and extended reality (XR), the services and applications of the Metaverse are gradually entering our lives.
However, the current development of the Metaverse provides users with services that are homogeneous with the user experience that the Internet has brought in the past, making them more like an extension of the Internet.
In addition, as a mobile application carrier for the Metaverse, it is also worth considering how vehicles with diverse onboard components can develop in synergy with the Metaverse.
In this article, we focus on the core of the Metaverse, namely user experience, and provide a road map from Metaverse to Metaverse vehicles (Metavehicles).
Specifically, we first elaborate on six features of the Metaverse from the perspective of user experience and propose a hierarchical framework for the Metaverse based on the evolutionary logic of the features. Under the guidance of this framework, we discuss the empowerment of onboard components of Metavehicles on the development of the Metaverse, and analyze the service experience that Metavehicles can bring to two types of users, namely drivers and passengers.
Finally, considering the differentiated development levels of Metaverse and autonomous driving, we further establish a hierarchical framework for Metavehicles from three aspects (i.e., enhance Metaverse, enhance driving experience, and enhance entertainment experience), providing an evolutionary path for the development of Metavehicles.
\end{abstract}
}


\maketitle

\IEEEdisplaynontitleabstractindextext

\renewcommand{\thefootnote}{}
\footnotetext{
Corresponding author: Nan Cheng, e-mail: dr.nan.cheng@ieee.org}

\section{Introduction}
\IEEEPARstart{T}{he} renaming of Facebook in October 2021 marked the new beginning of the Metaverse. Following this trend, as shown in Fig. 1, various Internet and vehicle companies around the world have also turned their attention to key technologies that enable diverse services and applications in the Metaverse.
For example, BMW has showcased their Metaverse product (i.e., BMW i Vision Dee) to the world at the Consumer Electronics Show in January 2023. The product can display information on the extended large-scale head-up display and fully mobilize different senses to create an immersive experience. In addition, through the design of headlights and closed BMW kidney grille, the vehicle can communicate with people and express emotions.

\begin{figure}[t!]
	\centering
	\includegraphics[width=1\linewidth]{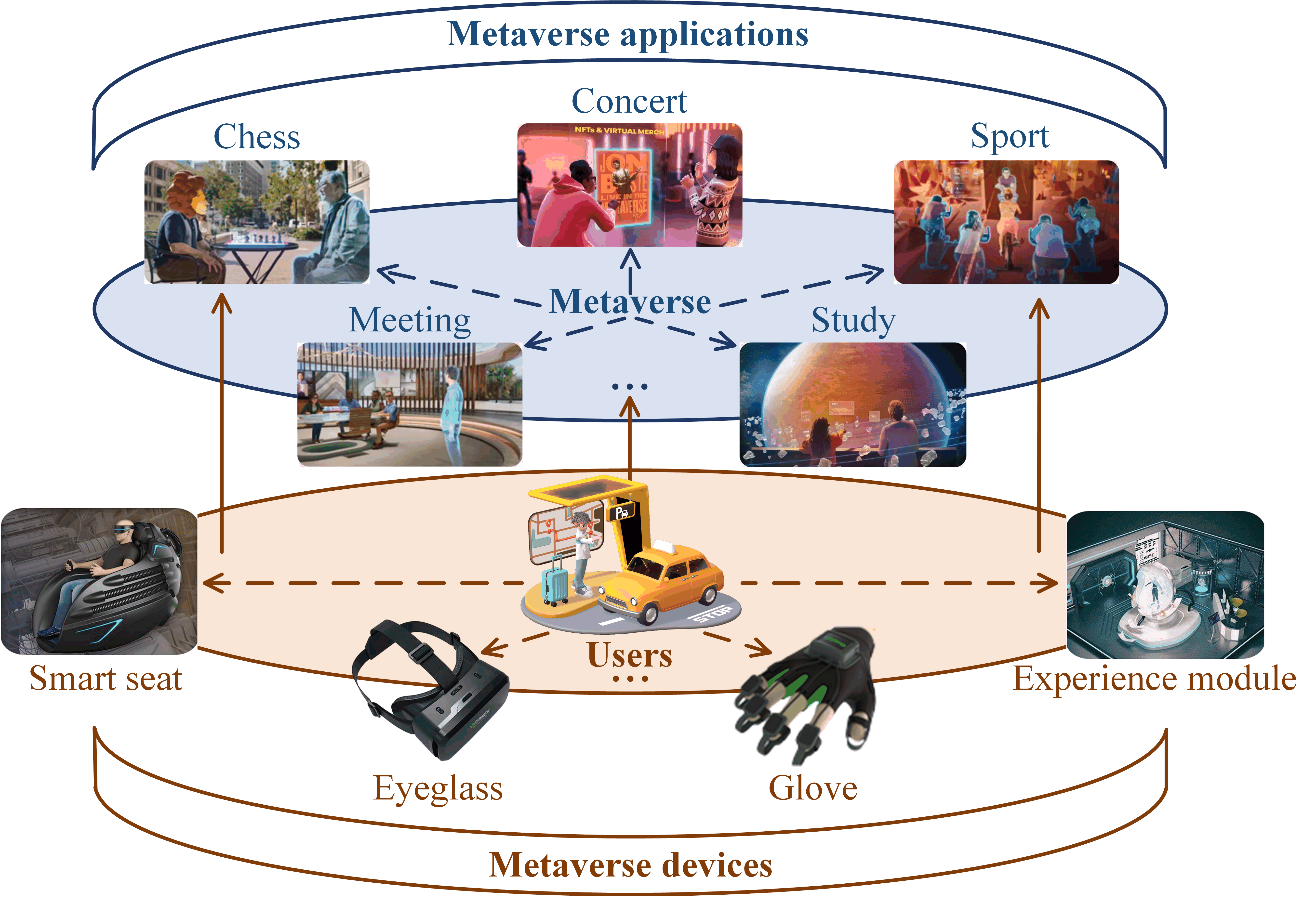}
	\caption{The devices and applications in the Metaverse.}
	\label{fig:1}
\end{figure}

Although the products of the Internet and vehicle companies aim to enable users to directly or indirectly experience the charm of the Metaverse, the current development of the Metaverse provides users with services that are homogeneous with the experience that the Internet can bring. Thus, it is more like an extension of the Internet from the perspective of user experience. In general, as a virtual world dominated by user behavior, the Metaverse should exhibit a distinct characteristic from the Internet to significantly improve user experience.

On the other hand, the intensive application of Metaverse elements such as virtual conferences, emotional interactions, and digital spaces in vehicle companies has made vehicle users not satisfied with traditional driving experiences, but more eager to intuitively experience the Metaverse in vehicles, such as immersive driving and virtual games. This promotes vehicles to adapt to the development of the Metaverse, meeting the user's progressive Metaverse experience, and promoting a safer, more sustainable, and more immersive driving experience.

In this article, we regard user experience as the core of the Metaverse and present a road map from Metaverse to Metaverse vehicles (Metavehicles).
We first elaborate on six features of Metaverse from the perspective of user experience. By considering the evolutionary logic of these six features, a hierarchical framework for the Metaverse is proposed to provide a basis for the discussion of the Metavehicles. Under the guidance of this framework, we discuss the empowerment of onboard components of Metavehicles on the development of the Metaverse, and divide users in Metavehicles into two types, namely drivers and passengers, to analyze the different Metaverse experiences. Finally, by considering the differentiated development
levels of Metaverse and autonomous driving, we establish a hierarchical framework for Metavehicles in terms of enhancing Metaverse, driving experience, and entertainment experience, thereby promoting the coordinated development of Metaverse and Metavehicles.

\section{Metaverse}
The Metaverse is essentially a virtual world that can provide users with experiences different from the real world. ROBLOX summarizes eight characteristics of the Metaverse from the perspective of social gaming, namely identity, friends, immersive, low friction, variety, anywhere, economy, and civility. Unlike ROBLOX, Matthew Ball describes the core attributes of the Metaverse as five aspects from the perspective of digital world, i.e., persistence, synchronicity, availability, economy, and interoperability\cite{MB}.

Obviously, different definitions and characteristics of the Metaverse can be given from different perspectives \cite{Du2023reth}. However, as a new technology that provides services to humanity, the Metaverse should establish a comprehensive user experience system at the beginning of its design. Therefore, in this article, we focus on the characteristics of the Metaverse in terms of user experience, as shown in Fig. 2, which are sequentially expressed as immersion, latency, real-time synchronization, identity, economic system, and the ultimate goal of the Metaverse, namely, to make people feel the digital civilization.

\begin{figure}[t!]
	\centering
	\includegraphics[width=1\linewidth]{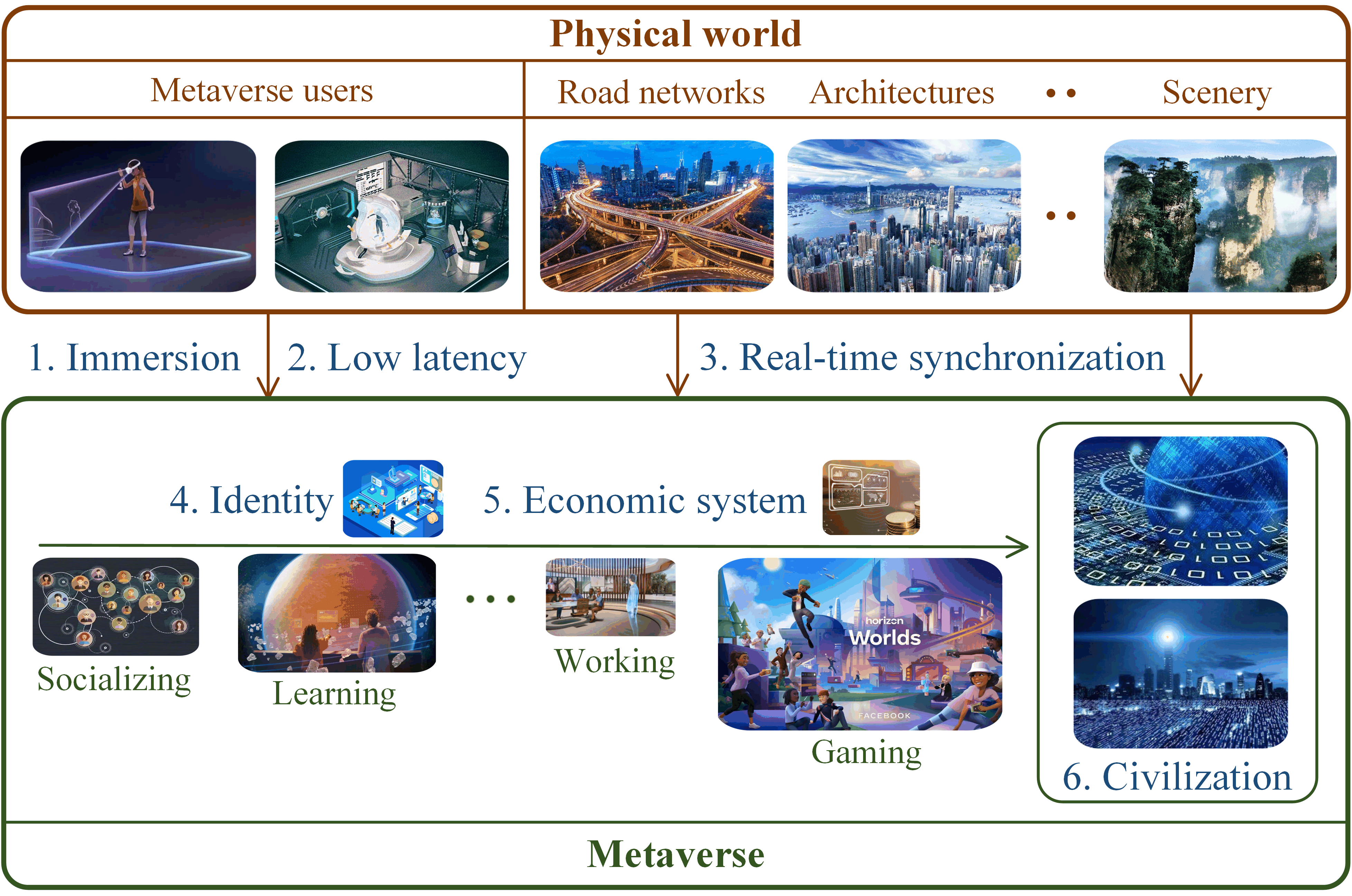}
	\caption{The features of the Metaverse in
terms of user experience.}
	\label{fig:2}
\end{figure}

\textbf{Immersion:} As the first step for users to experience the Metaverse, the way they connect to the Metaverse is the core factor that affects immersion. With current technological means, there are no obstacles for users to access the Metaverse through mobile or desktop devices. However, these access methods can easily distract people's attention and prevent them from fully immersing themselves in the Metaverse. Therefore, the Metaverse should provide users with more immersive access methods. By using digital senses such as virtual vision, virtual touch, virtual smell, etc., people are encouraged to believe that they are truly in the Metaverse\cite{Pan2022digital}. For example, creating realistic environments in a person's field of vision, providing virtual tactile feedback to people, and even using simulated odors to confuse their taste. The coupled development of sensations generated by multiple senses of the human body can enable users to connect their ears, nose, mouth, and even all body organs to the virtual world, providing a more comprehensive sense of immersion.

\textbf{Latency:} After accessing the Metaverse, users hope that the virtual world can provide timely feedback on their operations\cite{Zhou2023meta}. This requirement is reflected in the network latency including communication latency and computation latency. Powered by advanced 6G communication\cite{Asl2023meta} and edge computing technology\cite{Lim2022real}, the Metaverse can realize low latency services in diverse scenarios, helping users gain feedback experiences.

\textbf{Real-time synchronization:} After solving the delay problem, the first thing users can experience is the various elements in the Metaverse. In fact, many digital elements in the Metaverse are built on the basis of elements in the physical world. Through digital twin technology, real-time synchronization can be achieved between elements in the physical world and the underlying digital elements in the Metaverse. The Metaverse can then use these digital elements to update and iterate the state of the entire Metaverse, thereby more closely coupling the digital world and the physical world\cite{Liu2022meta,Han2023ady,Shi2023marl}.

\textbf{Identity:} In the Metaverse, users have the need to socialize, work, entertain, and learn across time and space. Driven by these requirements, users need to experience different identities in different scenarios. For example, users can maintain their original identity at work. At the same time, users can transform into medieval knights in entertainment services. Different needs lead to different scenarios and identities in the Metaverse. With the assistance of artificial intelligence, the Metaverse can adaptively generate and manage user identities, allowing users to seamlessly switch between different scenarios.

\textbf{Economic system:} The Metaverse needs an economic system as its driving force to support its operation. Through decentralized blockchain technology, the economic system in the Metaverse can possess the same credit as in the real economic system, thus ensuring the security of user identities and assets\cite{Fu2023asu}. The Metaverse economic system, including digital currency, digital goods, and digital markets, will involve all scenarios of the Metaverse. For example, users can obtain a certain share of digital currency through work in the Metaverse, which can be used to purchase digital products such as skins, non-fungible tokens, artworks, etc. in the digital market.

\textbf{Civilization:} The ultimate goal of the Metaverse is to showcase the brilliant digital civilization to users. With the increase and improvement of Metaverse applications and functions, users will develop unified rules in the Metaverse. Under the guidance of rules, families, languages, religions, cities, villages, and countries then can be generated in the Metaverse, ultimately evolving into the civilization of the Metaverse.

\section{Levels of Metaverse}
In this section, as shown in Fig. 3, we divide the evolution of the Metaverse into four levels based on user experience and technological development, providing a predictable path for the development of the Metaverse.

\begin{figure}[t!]
	\centering
	\includegraphics[width=1\linewidth]{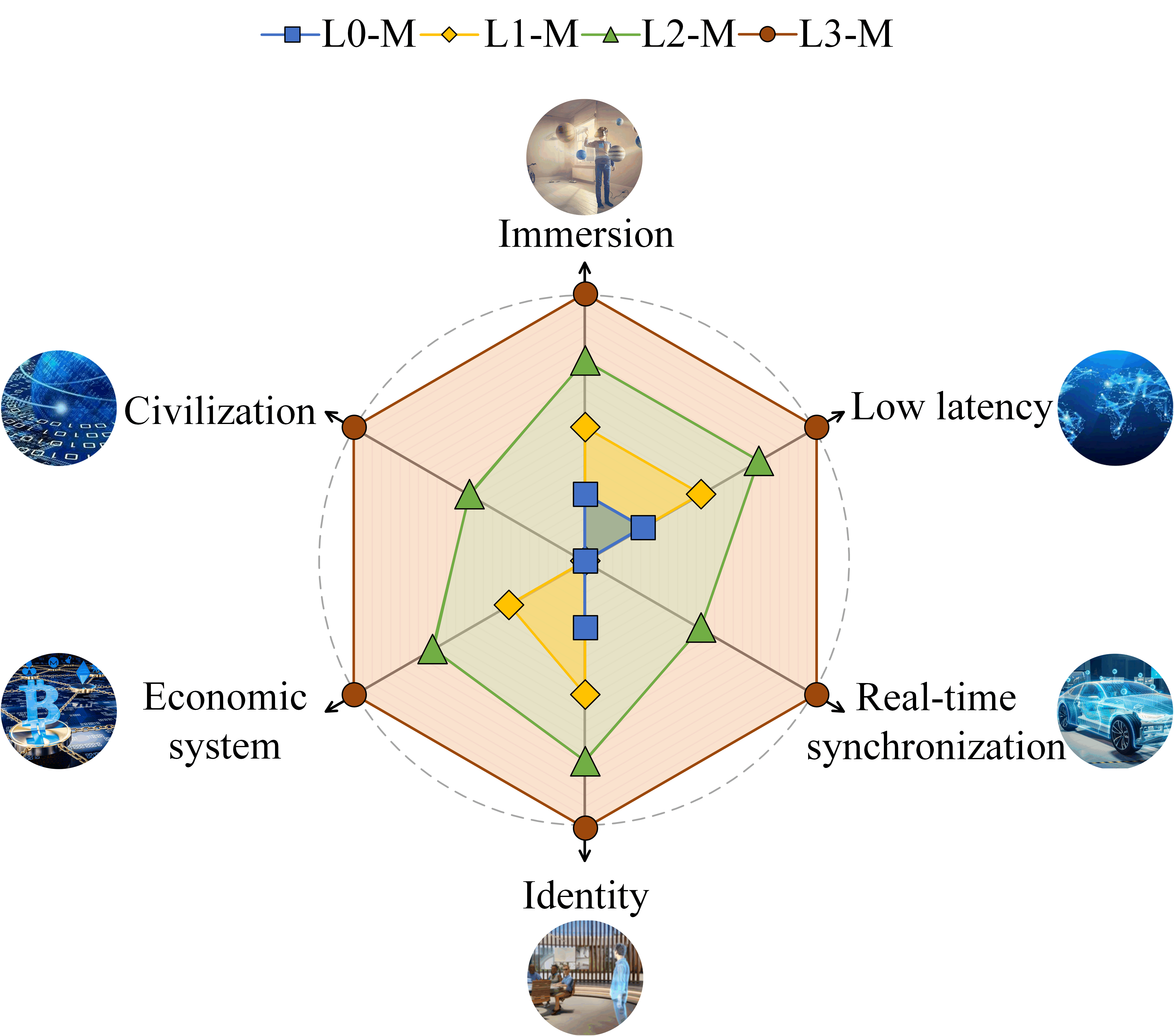}
	\caption{The levels of the Metaverse.}
	\label{fig:3}
\end{figure}

\subsection{L0-M}	
The L0-M refers to the development level of the Metaverse before 2021. At L0-M, users' entertainment, work, social, and learning needs are met by both the physical world and the Internet.
\begin{itemize}
	\item \textbf{Immersion:} Users can adopt a mouse or finger to click on the screen of a mobile or desktop device to enter the Metaverse. At this level, the sense of immersion is weak and susceptible to interference from external factors such as ambient light and sound.
	\item \textbf{Latency:} At L0-M, mainstream communication technologies such as 4G, wireless local area network, and wired communication are applied to link to the Metaverse. With these technologies, the delay varies depending on the specific communication method, resulting in significant changes in user experience.
	\item \textbf{Real-time synchronization:} At this level, digital twins are only linked to the real world through scattered text or images in a small part of industrial manufacturing. Therefore, it is difficult to create digital twins for all physical entities in the Metaverse.
	\item \textbf{Identity:} The Metaverse cannot fully take over users' work, social, learning, and entertainment scenarios. In addition, the switching of various scenes heavily relies on people's subjective operations. Therefore, there is a subjective lag in consciousness when switching from an old scene to a new one, making it difficult to quickly adapt to the new scene.
	\item \textbf{Economic system:} There is no stable economic system in the Metaverse. Even the blockchain tokens that have emerged in recent years have not been widely used like real currencies.
	\item \textbf{Civilization:} There is no concept of civilization at L0-M.
\end{itemize}

\subsection{L1-M: Creation of Virtual Elements}
At L1-M, the Metaverse is just beginning to develop. The elements in the network are not strictly synchronized with the real world, which means that there is more virtual content in the Metaverse. Content manufacturers often create and upload formed virtual content into the Metaverse for users to choose from. In addition, users can access the Metaverse not only through traditional mobile devices and computers, but also through wearing devices or gloves to allow a portion of the human body to enter the Metaverse and enjoy the Metaverse services.

\begin{itemize}
	\item \textbf{Immersion:} While compatible with L0-M's ability to enter the Metaverse, L1-M also has access methods such as headworn virtual reality (VR) devices and tactile gloves. Thus, the immersion can be improved compared to L0-M. However, the types of sensations that users receive from the Metaverse are singular.
	\item \textbf{Latency:} At this level, 5G has begun to be widely used, basically meeting the needs of users to access the Metaverse. However, the delay is still influenced by the scenario and time period, such as high service delay during hot periods.
	\item \textbf{Real-time synchronization:} At L1-M, the elements of the Metaverse are composed of two parts. The first part is the outdated content in the real world manually uploaded by the maintainer. The second part is to directly construct virtual elements in the Metaverse. In addition, event updates in the Metaverse mainly come from user actions, such as ROBLOX and other Metaverse game platforms.
	\item \textbf{Identity:} Users can generate different identities and complete identity switching in different scenarios, but there are fewer adaptation scenarios. The Metaverse has taken over some users' social and entertainment scenarios, while work, study, and other scenarios have not received enough attention.
	\item \textbf{Economic system:} Virtual economic systems have begun to emerge. Users can freely trade digital goods and assets in the Metaverse. However, the credit of the Metaverse economic system is weak and only partially recognized in fixed scenarios. Compared to real currencies, there is a lack of unified and stable credit.
	\item \textbf{Civilization:} Various modules related to civilization in the Metaverse, such as families, languages, religions, cities and villages, began to develop independently, but did not give birth to the concept of overall civilization.
\end{itemize}

\subsection{L2-M: Extension of Reality}

The iconic milestone of L2-M is the realization of real-time synchronization between real and digital elements, that is, access to reality, so that the Metaverse constructed on digital elements can be seen as an extension of the real world.
\begin{itemize}
	\item \textbf{Immersion:} Compared to L1-M, L2-M's Metaverse can provide users with a better sense of immersion. Through full body wearable devices or enclosed experience silos, users can connect all their senses to the virtual world, truly immersing themselves in the Metaverse and obtaining various interactive sensations.
	\item \textbf{Latency:} At L2-M, 6G begins to develop to meet the needs of users accessing the Metaverse. Compared to 5G, the initially developed 6G has significant advantages in suburban and hot periods.
	\item \textbf{Real-time synchronization:}  L2-M achieves real-time one-way synchronization between real elements and digital elements. Users can experience real world elements or create virtual elements based on real world elements without worrying about these digital elements being disconnected from real life.
	\item \textbf{Identity:} The Metaverse can basically take over common scenarios such as social, entertainment, work, and learning in users' lives. Furthermore, it can achieve seamless switching of users' identities in various virtual scenarios.
	\item \textbf{Economic system:} The virtual economic system is gradually stabilizing. Users can not only freely trade digital goods in the virtual world, but also use virtual currency to trade real goods. This means that the purchasing power and credit of virtual currency are not limited to the Metaverse, but even extend to the physical world.
	\item \textbf{Civilization:} Various civilization related modules in the Metaverse have begins to develop in a cross disciplinary manner, where users can live, grow, and even change the evolution of digital civilization in the virtual world.
\end{itemize}
\subsection{L3-M: Combination of Virtual and Reality}

After achieving real-time synchronization of real elements, the next development direction of the Metaverse is to bring characters, scenes, and services from the virtual world into the real world. By utilizing new technologies such as holographic projection and augmented reality, users do not need to access the Metaverse and can also experience the services of the Metaverse in the real world. Therefore, the characteristic of this level is that the virtual of the Metaverse and the reality of the world begin to interweave and couple, achieving the unity of virtual and reality.
\begin{itemize}
	\item \textbf{Immersion:} At L3-M, users can not only choose to access the Metaverse through external devices, but also experience virtual scenes and services from anywhere in the real world. The experiences include seeing virtual characters, hearing virtual audio, smelling the fragrance of virtual food, and so on.
	\item \textbf{Latency:} With the large-scale application of 6G, the Metaverse can be accessed anytime and anywhere and the virtual elements in the Metaverse can be displayed in the real world in real time.
	\item \textbf{Real-time synchronization:}  At this level, the real world and Metaverse can achieve bidirectional real-time synchronization. This means that elements in the real world can achieve real-time synchronization with corresponding digital elements in the Metaverse. At the same time, the original elements in the virtual world will also be synchronized with their physical representations in the physical world.
	\item \textbf{Identity:} The deep integration of the virtual world and the real world allows users to easily adjust various application scenarios and switch identities to adapt to all scenarios.
	\item \textbf{Economic system:} The economic system in the Metaverse is deeply integrated with the real economic system, giving birth to a new unified economic system that is universal between the real world and virtual networks. In this system, a unified currency can circulate in both the virtual and real worlds for users to consume and invest.
	\item \textbf{Civilization:} The dispersed modules of digital civilization develop in a unified and coordinated manner, and gradually integrate with real civilization to form a new Earth civilization. In the ultimate civilization, digital life, digital races, and digital nations become part of human life, and more and more individuals choose to trade, create, and live in the digital world of the Metaverse.
\end{itemize}

\section{Metavehicles}
Connected and automated vehicles, as mobile platforms equipped with various sensing and computing devices, can be used to enhance the Metaverse\cite{Zhang2022para,zhou2023vetaverse,Ren2022qua}. On the other hand, immersive Metaverse applications can improve the entertainment and driving experience of drivers and passengers on the go. Therefore, in this section, we explain the role of vehicles in the Metaverse based on the main onboard components of connected and automated vehicles shown in Fig. 4.

\begin{figure}[t!]
	\centering
	\includegraphics[width=1\linewidth]{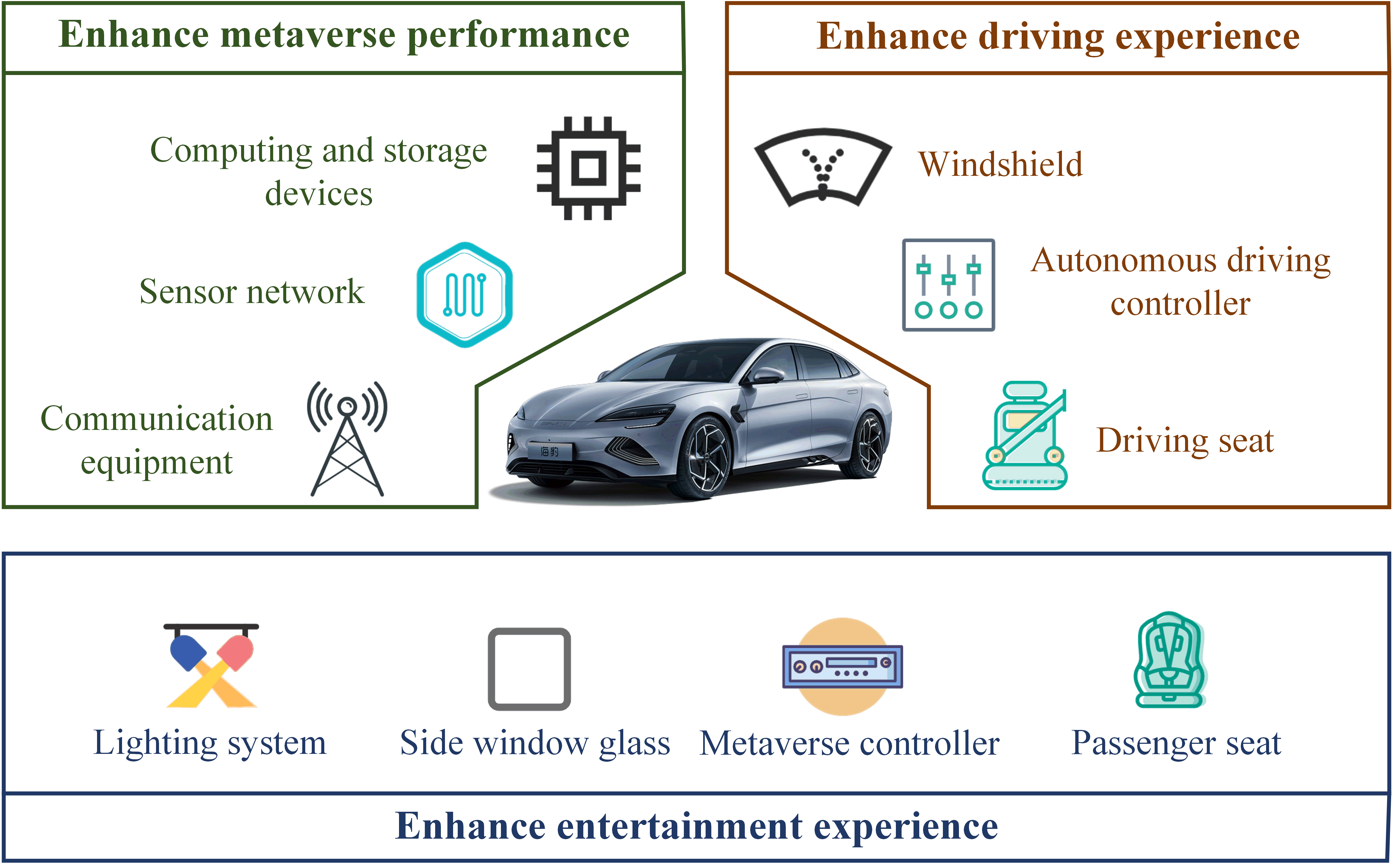}
	\caption{The onboard components of the Metavehicles.}
	\label{fig:4}
\end{figure}

\subsection{Enhance Metaverse Performance}
\textbf{Sensor network:} The sensor network of a vehicle can be divided into internal sensor network and external sensor network. The internal sensor network includes position sensors, temperature sensors, acceleration sensors, angular velocity sensors, etc., can be used to perceive vehicle status data and upload the collected data to the Metaverse. In addition, the collected data can be used to synchronize the states of corresponding virtual vehicles in the Metaverse\cite{xu2023generative} so that vehicle companies can analyze the data and provide customized optimization.

\textbf{Storage and computing devices:} In addition to executing driving tasks, meeting entertainment needs, and storing vehicle data, vehicle storage and computing devices can also provide computing, storage, and offloading services for the Metaverse by using idle resources, thereby reducing the load on the master server.	

\textbf{Communication equipment:} The communication equipment of the vehicle can not only be connected to the Metaverse to provide collected data and offloaded tasks, but also has the ability to enable passengers and drivers in the vehicle to access the Metaverse with low latency.
\subsection{Enhance Driving Experience}
\textbf{Autonomous driving controller:} Autonomous driving controller is usually used to integrate heterogeneous data collected by internal and external sensors and make driving decisions, such as adjusting the driving posture of the vehicle and planning the driving route. In the Metaverse, autonomous driving controller can also use communication devices to connect to the Metaverse to obtain more road data or use the computing resources of the Metaverse to complete complex driving decisions.

\textbf{Driving seat:} The immersive driver's seat can capture the driver's emotions and states, and upload these data to the Metaverse for analysis. Subsequently, feedback such as vibration, sound, and odor can be provided to enhance the driver's attention. Specifically, in the event of an impending accident, other users or artificial intelligence from the Metaverse are allowed to take over the driver's operations to ensure driving safety.

\textbf{Windshield:} The front windshield of a vehicle can be considered as a screen for holographic projection. Specifically, the windshield can capture the driver's line of sight and use holographic projection to display the vehicle's status in front of the driver, including the vehicle's speed, energy consumption status, and the driving environment in front of and behind the vehicle. In addition, by overlaying virtual reality technology into the driving environment, local computing devices of the vehicle or the Metaverse can provide accurate driving suggestions, thereby avoiding frequent eye switches and enhancing driving safety.

\subsection{Enhance Entertainment Experience}
\textbf{Metaverse entertainment controller:} Metaverse entertainment controller can integrate vehicle computing, storage, and communication resources, connect users to the Metaverse, and enable users to freely switch between different Metaverse scenes. In addition, the entertainment controller can integrate the lighting, audio, screen, seats, etc. in the passenger area of the vehicle, creating a closed environment and providing users with high immersion Metaverse services. Finally, the entertainment controller will also maintain the virtual vehicle corresponding to the Metavehicle and integrate multiple types of sensor data to provide virtual services to virtual passengers in the Metaverse.

\textbf{Passenger seat:} The passenger seat, which is deeply integrated with the Metaverse, can provide real-time feedback on passengers' real experiences in the virtual world, including vibration, touch, hearing, and smell, thereby increasing passengers' immersion in the Metaverse. At the same time, the passenger seat can also capture information such as passengers' micro expressions, emotions, and states, thus creating customized Metaverse scenes for passengers\cite{Li2023intell}.

\textbf{Side window glass:} By cooperating with the passenger seat, the side window glass can be connected to the Metaverse to display the user's scene, achieving seamless switching between virtual and reality. On the other hand, the side window glass can project scenes from the Metaverse into the interior of the vehicle, thereby promoting users' experience in the combination of virtual and real scenes.

\textbf{Lighting system:} The lighting system can assist in implementing holographic projection to present digital elements in the Metaverse in a three-dimensional manner in the real world. In this way, passengers do not need to be connected to the Metaverse and can also experience the sense of coupling between virtual and reality. In addition, the lighting system can dynamically change with the changes in application scenarios in the Metaverse, promoting the coordination and unity of the environment inside the Metavehicle and the Metaverse environment.

\section{Levels of Metavehicles}
In the current transportation system, vehicles are more used as a means of transportation rather than sensing, computing, and entertainment platforms. Although WayRay produced the vehicle, Holograktor, that can be linked to a virtual world$^1$\footnote{$^1$https://holograktor.com/}, the development of vehicles in the market has not fully adapted to the needs of the Metaverse. With the gradual development of the Metaverse, how vehicles evolve and develop in synergy with the Metaverse has become a problem. Considering that autonomous vehicles differentiate the autonomous driving levels based on the driver's driving experience, as shown in Fig. 5, we refer to the development of autonomous driving and the classification of the Metaverse in Section III to intuitively distinguish different levels of Metavehicles based on the experience in the Metaverse. Obviously, it can be foreseen that for L5 autonomous vehicles the corresponding Metavehicles will eliminate the modules for the driver experience. This is because all driving operations of the vehicle will be taken over by the autonomous driving controller, and only the entertainment module for passengers can be seen inside the Metavehicle.

\begin{figure}[t!]
	\centering
	\includegraphics[width=1\linewidth]{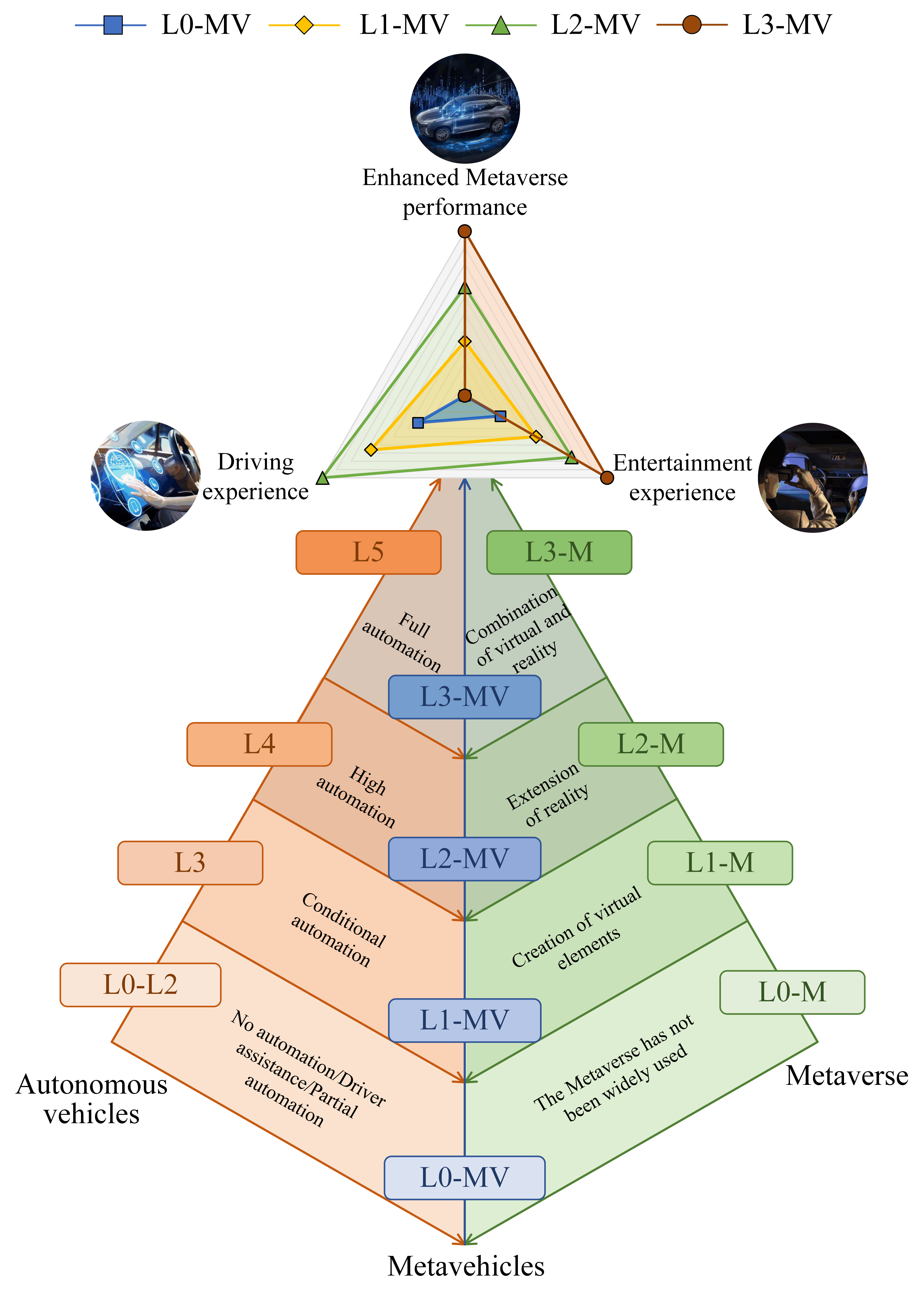}
	\caption{The levels of the Metavehicles.}
	\label{fig:5}
\end{figure}

\subsection{L0-MV}
The Metavehicles at L0-MV, like vehicles that are now widely used (i.e., L0-L2), were not manufactured with the consideration of Metaverse factors.
\begin{itemize}
	\item \textbf{Enhance Metaverse performance:} None. Because at this time, the Metaverse was not born.
	\item \textbf{Entertainment experience:} Passengers can access the Internet for entertainment through flat screens and audio devices in the Metavehicle. But compared to using screens and audio systems, passengers are more willing to use their own mobile devices, such as computers, phones, etc., to complete services.
	\item \textbf{Driving experience:} The vehicle can achieve partial autonomous driving to slightly enhance the driver's driving experience. For example, in specific scenarios, it can achieve automatic steering, acceleration, and braking.
\end{itemize}

\subsection{L1-MV: Access Tool}
At L1-MV, Metavehicles (i.e., L3) are mainly manufactured to adapt to the L1-M, meaning that Metavehicles are more commonly used as access tools for users to experience virtual worlds in the Metaverse.
\begin{itemize}
	\item \textbf{Enhance Metaverse performance:} Weak. Due to the lack of real-time synchronization between the Metaverse and physical Metavehicles at this level, Metavehicles can only sell their data to content manufacturers to create corresponding virtual content in the Metaverse. In addition, there are fewer applications in the Metaverse at L1-M, and the computational and storage resources consumed can be met by cloud servers. Consequently, Metavehicles do not need to provide their idle resources to facilitate the Metaverse.
	\item \textbf{Entertainment experience:} The preliminary development of passenger seats, combined with headworn VR devices, tactile gloves, etc., allows passengers to access the Metaverse. In addition, vibration, tactile, auditory, and odor feedback can be provided to convey a virtual world experience and provide passengers with an immersive experience. Besides, the side window glass can also display passengers' exploration journey in the Metaverse.
	\item \textbf{Driving experience:} The Metavehicle can achieve conditional autonomous driving, but the driver still needs to control the Metavehicle. At this level, the development of the driver's seat can enhance the driver's focus through feedback information. Additionally, the windshield displays the environment and status with the driver's line of sight, avoiding frequent eye switching while increasing driving safety.
\end{itemize}

\subsection{L2-MV: Comprehensive Service Equipment}
The L2-MV is fully adapted to the development of the L2-M. At this level, Metavehicles are not limited to being a simple access tool, but rather serve as a comprehensive service equipment, providing drivers and passengers with more realistic virtual experiences. At the same time, the further development of autonomous driving (L4) can enhance the driver's driving experience with the assistance of the Metaverse.
\begin{itemize}
	\item \textbf{Enhance Metaverse performance:} Moderate. The L2-M achieves real-time data synchronization between virtual and reality. Thus, the sensor networks of Metavehicles can collect vehicle status data, environmental data, etc., and then upload them to the Metaverse to provide a foundation for experiencing Metaverse services.
	\item \textbf{Entertainment experience:} The functions of the passenger seat are more diverse and practical. It can not only connect to the Metaverse and provide real-time feedback on feelings, but also capture passengers' micro expressions, emotions, and states, and synchronize them in real-time to the Metaverse. With these data, the Metaverse can provide personalized services and customized scenario suggestions for passengers.
	\item \textbf{Driving experience:}  The Metavehicle achieves highly autonomous driving and can independently complete all driving operations in most scenarios without the need for driver intervention. The development of driver's seat has reached a stage of large-scale commercial use, which can be integrated into the Metaverse to capture the driver's emotions and states, and upload these data to the Metaverse for analysis. Subsequently, feedback on vibrations, sounds, and odors can be used to enhance the driver's focus. In addition, the windshield of the Metavehicle can display not only environmental information and vehicle status, but also driving suggestions provided through the Metavehicle's local computing device or the Metaverse. By mixing virtual elements and overlaying them in the real environment, it can assist drivers in making more effective driving decisions.
\end{itemize}

\subsection{L3-MV: Immersive Experience Platform}
The L3-MV has achieved fully autonomous driving (L5). Therefore, there is no need to add additional driving experience modules in the Metavehicle, and more space can be released to enhance the vehicle's computing, storage, and communication capabilities. On the other hand, due to all driving operations are provided by physical devices or artificial intelligence in the Metaverse, the Metavehicle itself has fully evolved into an immersive experience platform for efficient mixing of virtual and reality.
\begin{itemize}
	\item \textbf{Enhance Metaverse performance:} Strong. In addition to achieving real-time synchronization, the additional resources of Metavehicles can also provide computing, storage, and offloading services. Therefore, the quality of the Metaverse can be improved by reducing latency and the load on the main server of the Metaverse.
	\item \textbf{Entertainment experience:} Except for the passenger seat and side window glass, the interior space of the vehicle can be fully utilized to enhance the user's entertainment experience. To adapt to the development of the L3-M, the holographic display systems will be added to the Metavehicle, integrating components such as lighting, seats, and side window glass. By adapting the holographic display system to scenes in the Metaverse, the digital elements in the virtual world can be presented in a three-dimensional manner in the real world, allowing passengers to experience the interweaving of virtual and reality.
	\item \textbf{Driving experience:}  Due to the highly developed autonomous driving technology, Metavehicles can efficiently and safely complete all driving operations in all scenarios with the assistance of artificial intelligence. Therefore, there is no need to enhance the driver's driving experience at this level.
\end{itemize}
\section{Conclusion}\label{sec:conclusion}

In this article, we have provided the road map from Metaverse to Metavehicles by paying attention to user experience. Specifically, we have summarized six characteristics of the Metaverse from the user's perspective. By comprehensively considering these six features and the evolution logic of the Metaverse, we have proposed a hierarchical framework for developing the Metaverse. Subsequently, by combining the hierarchical framework of the Metaverse with the onboard components of vehicles, we have discussed how vehicles closely coupled with the Metaverse to develop into Metavehicles. Finally, by integrating the different development levels of Metaverse and autonomous driving, a hierarchical framework for Metavehicles has been established, providing an achievable path for the development of Metavehicles.

Although the Metaverse has not yet become widespread, it can be foreseen from the two hierarchical frameworks presented in this article that the Metaverse will rapidly improve people's life experiences in the near future. In addition, Metavehicles will evolve in synergy with the Metaverse and ultimately completely change people's way of travel.
\bibliographystyle{IEEETran}
\bibliography{ref}

\begin{thebibliography}{10}
\providecommand{\url}[1]{#1}
\csname url@samestyle\endcsname
\providecommand{\newblock}{\relax}
\providecommand{\bibinfo}[2]{#2}
\providecommand{\BIBentrySTDinterwordspacing}{\spaceskip=0pt\relax}
\providecommand{\BIBentryALTinterwordstretchfactor}{4}
\providecommand{\BIBentryALTinterwordspacing}{\spaceskip=\fontdimen2\font plus
\BIBentryALTinterwordstretchfactor\fontdimen3\font minus
  \fontdimen4\font\relax}
\providecommand{\BIBforeignlanguage}[2]{{%
\expandafter\ifx\csname l@#1\endcsname\relax
\typeout{** WARNING: IEEEtran.bst: No hyphenation pattern has been}%
\typeout{** loaded for the language `#1'. Using the pattern for}%
\typeout{** the default language instead.}%
\else
\language=\csname l@#1\endcsname
\fi
#2}}
\providecommand{\BIBdecl}{\relax}
\BIBdecl

\bibitem{MB}
\BIBentryALTinterwordspacing
M.~Ball, ``The metaverse: What it is, where to find it, and who will build
  it.'' Jan. 2020. [Online]. Available:
  \url{https://www.matthewball.vc/all/themetaverse}
\BIBentrySTDinterwordspacing

\bibitem{Du2023reth}
H.~Du, B.~Ma, D.~Niyato, J.~Kang, Z.~Xiong, and Z.~Yang, ``Rethinking quality
  of experience for metaverse services: A consumer-based economics
  perspective,'' \emph{IEEE Network}, pp. 1--8, Early Access 2023.

\bibitem{Pan2022digital}
D.~Panagiotakopoulos, G.~Marentakis, R.~Metzitakos, I.~Deliyannis, and
  F.~Dedes, ``Digital scent technology: Toward the internet of senses and the
  metaverse,'' \emph{IT Professional}, vol.~24, no.~3, pp. 52--59, May-June
  2022.

\bibitem{Zhou2023meta}
Y.~Zhou, H.~Huang, S.~Yuan, H.~Zou, L.~Xie, and J.~Yang, ``Metafi++:
  Wifi-enabled transformer-based human pose estimation for metaverse avatar
  simulation,'' \emph{IEEE Internet of Things Journal}, pp. 1--1, Early Access
  2023.

\bibitem{Asl2023meta}
A.~M. Aslam, R.~Chaudhary, A.~Bhardwaj, I.~Budhiraja, N.~Kumar, and
  S.~Zeadally, ``Metaverse for 6g and beyond: The next revolution and
  deployment challenges,'' \emph{IEEE Internet of Things Magazine}, vol.~6,
  no.~1, pp. 32--39, Mar. 2023.

\bibitem{Lim2022real}
W.~Y.~B. Lim, Z.~Xiong, D.~Niyato, X.~Cao, C.~Miao, S.~Sun, and Q.~Yang,
  ``Realizing the metaverse with edge intelligence: A match made in heaven,''
  \emph{IEEE Wireless Communications}, pp. 1--9, Early Access 2022.

\bibitem{Liu2022meta}
Y.~Liu, Y.~Shen, C.~Guo, Y.~Tian, X.~Wang, Y.~Zhu, and F.-Y. Wang,
  ``Metasensing in metaverses: See there, be there, and know there,''
  \emph{IEEE Intelligent Systems}, vol.~37, no.~6, pp. 7--12, Nov.-Dec. 2022.

\bibitem{Han2023ady}
Y.~Han, D.~Niyato, C.~Leung, D.~I. Kim, K.~Zhu, S.~Feng, X.~Shen, and C.~Miao,
  ``A dynamic hierarchical framework for iot-assisted digital twin
  synchronization in the metaverse,'' \emph{IEEE Internet of Things Journal},
  vol.~10, no.~1, pp. 268--284, Jan. 2023.

\bibitem{Shi2023marl}
H.~Shi, G.~Liu, K.~Zhang, Z.~Zhou, and J.~Wang, ``Marl sim2real transfer:
  Merging physical reality with digital virtuality in metaverse,'' \emph{IEEE
  Transactions on Systems, Man, and Cybernetics: Systems}, vol.~53, no.~4, pp.
  2107--2117, Apr. 2023.

\bibitem{Fu2023asu}
Y.~Fu, C.~Li, F.~R. Yu, T.~H. Luan, P.~Zhao, and S.~Liu, ``A survey of
  blockchain and intelligent networking for the metaverse,'' \emph{IEEE
  Internet of Things Journal}, vol.~10, no.~4, pp. 3587--3610, Feb. 2023.

\bibitem{Zhang2022para}
H.~Zhang, G.~Luo, Y.~Li, and F.-Y. Wang, ``Parallel vision for intelligent
  transportation systems in metaverse: Challenges, solutions, and potential
  applications,'' \emph{IEEE Transactions on Systems, Man, and Cybernetics:
  Systems}, pp. 1--14, Early Access 2022.

\bibitem{zhou2023vetaverse}
\BIBentryALTinterwordspacing
P.~Zhou, J.~Zhu, Y.~Wang \emph{et~al.}, ``Vetaverse: A survey on the
  intersection of metaverse, vehicles, and transportation systems,'' 2023.
  [Online]. Available: \url{https://arxiv.org/abs/2210.15109}
\BIBentrySTDinterwordspacing

\bibitem{Ren2022qua}
Y.~Ren, R.~Xie, F.~R. Yu, T.~Huang, and Y.~Liu, ``Quantum collective learning
  and many-to-many matching game in the metaverse for connected and autonomous
  vehicles,'' \emph{IEEE Transactions on Vehicular Technology}, vol.~71,
  no.~11, pp. 12\,128--12\,139, Nov. 2022.

\bibitem{xu2023generative}
\BIBentryALTinterwordspacing
M.~Xu, D.~Niyato, J.~Chen, H.~Zhang, J.~Kang, Z.~Xiong, S.~Mao, and Z.~Han,
  ``Generative ai-empowered simulation for autonomous driving in vehicular
  mixed reality metaverses,'' 2023. [Online]. Available:
  \url{https://arxiv.org/abs/2302.08418}
\BIBentrySTDinterwordspacing

\bibitem{Li2023intell}
W.~Li, L.~Wu, C.~Wang, J.~Xue, W.~Hu, S.~Li, G.~Guo, and D.~Cao, ``Intelligent
  cockpit for intelligent vehicle in metaverse: A case study of empathetic
  auditory regulation of human emotion,'' \emph{IEEE Transactions on Systems,
  Man, and Cybernetics: Systems}, vol.~53, no.~4, pp. 2173--2187, Apr. 2023.

\end{thebibliography}

\end{document}